\documentclass[%
 reprint,
showpacs,preprintnumbers,
 amsmath,amssymb,
 aps,
prb,
]{revtex4-1}

\usepackage{graphicx}% Include figure files
\usepackage{dcolumn}% Align table columns on decimal point
\usepackage{bm}% bold math
\usepackage{color}
\usepackage{ulem}
\usepackage{amssymb, amsmath}

\begin{document}

\preprint{APS/123-QED}

\title{Shape of magnetic domain walls formed by coupling to mobile charges}

\author{Ryo Ozawa,$^1$ Satoru Hayami,$^{2}$ Kipton Barros,$^{3}$ and Yukitoshi Motome$^1$}
\affiliation{%
$^1$Department of Applied Physics, University of Tokyo, Tokyo 113-8656, Japan\\
$^2$Department of Physics, Hokkaido University, Sapporo 060-0810, Japan\\
$^3$Theoretical Division and CNLS, Los Alamos National Laboratory, Los Alamos, New Mexico 87545, U.S.A.
 }%

\date{\today}% It is always \today, today,
 
\begin{abstract}
Magnetic domain walls, which are crucially important in both fundamental physics and technical applications, 
often have a preference in their form due to many different origins, such as the crystalline shape, lattice symmetry, and magnetic anisotropy.
We theoretically investigate 
yet another origin stemming from the coupling to mobile charges in itinerant magnets.
Performing a large-scale numerical simulation  in a minimal model for itinerant magnets, i.e., the Kondo lattice model with classical localized spins, 
 we show that the shape of magnetic domain walls 
depends on the electronic band structure and electron filling. 
While N\'eel and 120$^\circ$ antiferromagnetic states
do not show a strong preference in the shape of domain walls,
noncoplanar spin states with scalar chiral ordering have distinct directional preferences of the domain walls depending on the electron filling. 
We find that the directional preference is rationalized by
the wave-number dependence of the effective magnetic interactions 
induced by the mobile charges, which are set by the band structure and electron filling. 
We also observe that, in the noncoplanar chiral states, an electric current is induced along the domain walls 
owing to the spin Berry phase mechanism, with very different spatial distributions depending on whether the bulk state is metallic or insulating. 
\end{abstract}
\pacs{71.10.Fd, 71.27.+a, 75.10.-b}
\maketitle

\section{Introduction}
\label{sec:Introduction}

Magnetic domain walls (DWs) have been an important issue in both fundamental physics of magnetism and applications to magnetic devices.
Besides the phenomenological understanding, it is not easy to establish the microscopic theory for the formation of DWs, 
as it is basically a nonequilibrium phenomenon with spatial inhomogeneity, often ranging from nano to micrometer scales. 
Magnetic DWs have been studied 
mainly in classical spin models with neglecting mobile charges, by considering the effects of, e.g., crystalline shape, lattice symmetry, and magnetic anisotropy\cite{RevModPhys.21.541, chikazumi2009physics, bozorth1937directional}.
Spatial modulations of spin textures have been clarified near the DWs, 
such as the so-called Bloch, N\'eel, and cross-tie walls. 

Recently, the magnetic DWs have gained renewed interest 
through the studies toward functional devices based on, 
e.g., the giant magnetoresistive effect\cite{PhysRevB.51.9855, PhysRevLett.77.1580}, spintronics\cite{PhysRevLett.78.3773, PhysRevLett.79.5110, Tatara2008213}, and multiferroics\cite{pvrivratska2002magnetic,seidel2009conduction}. 
An intriguing issue 
is the current-induced DW motion. 
The controllability of DWs by an electric current 
was experimentally demonstrated in itinerant ferromagnets\cite{PhysRevLett.92.077205, yamanouchi2004current}.
The microscopic mechanism was theoretically discussed by
taking into consideration the interplay between spin and charge degrees of freedom of electrons\cite{SLONCZEWSKI1996L1, PhysRevB.83.174447, PhysRevLett.107.136804, PhysRevB.54.9353, PhysRevLett.95.026601, Tatara2008213, doi:10.7566/JPSJ.84.083701}. 
Another issue discussed very recently is the DWs in topological states of matter.  
An example has been discussed in the peculiar magnetically ordered state 
called all-in all-out type in pyrochlore oxides. 
For instance, in Cd$_2$Os$_2$O$_7$, the domain formation was observed 
using resonant x-ray diffraction, 
and controllability of the domain structures by a magnetic field-cooling procedure was demonstrated\cite{PhysRevLett.114.147205}.
Meanwhile, for iridium pyrochlore oxides, 
peculiar electronic states 
were theoretically predicted, such as the Weyl semimetal~\cite{PhysRevB.83.205101} and 
peculiar metallic DWs\cite{PhysRevX.4.021035, PhysRevB.93.195146}.
Experimentally, a hysteresis  
was observed in the magnetoresistance for Nd$_2$Ir$_2$O$_7$, and its relation to the metallic DWs was discussed\cite{JPSJ.82.023706, PhysRevB.89.075127, PhysRevB.92.121110}.
Recently, the real-space image of the metallic DWs was observed by microwave impedance microscopy~\cite{Ma538}.

In these DW problems, 
an interesting fundamental issue is how the shape of DWs is determined in such itinerant magnets.
In most of the previous studies, the presence and the shape of DWs were introduced `by hand', 
and the resultant electronic and magnetic properties were studied.
For a microscopic understanding of the DW physics, 
it is crucial to study the DW formation by treating the spin and charge (and possibly, orbital) degrees of freedom on an equal footing, 
without posing any assumption on the patterns of DWs.
It is also crucial to incorporate spatial, thermal, and quantum fluctuations for 
further development, including the creation, annihilation, deformation, and movement of DWs. 
Amongst others, it is strongly desired to enable unbiased calculations of such complicated systems, 
in spatially large systems ranging up to nano or micrometer scales.

In this paper, as a first step toward such further understanding of DWs in itinerant magnets, 
we theoretically study the formation of DWs by large-scale numerical simulation including  
the interplay between spin and charge degrees of freedom. 
Our purpose is to clarify what types of magnetic domains are formed through the 
interplay. 
In particular, our focus is on how the shape of the DWs is determined 
for the systems with different electronic and lattice structures. 
We also study the electronic properties near the magnetic DWs. 

For this purpose, we consider a minimal model for the spin-charge coupled systems, 
the Kondo lattice model, which describes mobile charges coupled to classical localized spins. 
To enable large-scale simulation, we employ a numerical technique based on the Langevin dynamics (LD) and kernel polynomial method (KPM)\cite{PhysRevB.88.235101}.  
In Ref.~\onlinecite{PhysRevB.88.235101},
different shapes of  
the DWs were reported for different 
electron fillings even for the same magnetic structure, 
while the origin was not referred. 
Stimulated by the observation, we here study the DW issue in the Kondo lattice model in 
a systematic way, using a modified version of the KPM-LD method by applying stochastic Landau-Lifshitz (SLL) dynamics, away from the overdamped limit.

In the present study, we analyze the formation of magnetic domains 
by performing the modified KPM-LD simulation with a sudden 
quench from high-temperature limit to zero temperature. 
We study several different magnetically ordered states appearing in the Kondo lattice models on square and triangular lattices while varying the electron filling.
We find that, for the collinear N\'eel state on the 
square lattice and coplanar $120^{\circ}$ state on the triangular lattice,
the DWs do not show a strong preference in their directions in real space.
On the other hand, for the noncoplanar spin states with three ordering vectors (triple-$Q$) on 
the triangular lattice near 1/4 and 3/4 fillings\cite{PhysRevLett.101.156402, JPSJ.79.083711}, 
we demonstrate that the DWs show very distinct shapes depending on the electron filling, as 
 observed in the previous study\cite{PhysRevB.88.235101}.
We also show that spontaneous electric current, which emerges 
through the spin Berry phase mechanism\cite{PhysRevLett.69.3232,PhysRevLett.83.3737}, 
flows along the DWs 
in a different manner
between the two cases
despite the same magnetic ground states. 
We show that the distinct directional preference of DWs is  
brought by 
the momentum-dependence of effective spin-spin interactions mediated by mobile charges.

The rest of the paper is organized as follows.
In Sec.~\ref{sec:Model and method}, we introduce the model 
and numerical method that we use in the present study. 
After introducing the Hamiltonian of the Kondo lattice model in Sec.~\ref{sec:KLM}, 
we discuss effective interactions between the localized spins 
in Sec.~\ref{sec:RKKY}. 
We also introduce the method of
KPM-LD simulation 
and its extension in Sec.~\ref{sec:KPM-LD}, and a numerical method to evaluate the expectation value of electronic properties in Sec.~\ref{sec:localKPM}. 
In Sec.~\ref{sec:results}, we 
present the numerical results. 
We show the DW formation 
in the collinear N\'eel state on the square lattice (Sec.~\ref{sec:AFM}), 
the coplanar $120^\circ$ state on the triangular lattice (Sec.~\ref{sec:120}),  
 and noncoplanar  triple-$Q$ states on the triangular lattice (Sec.~\ref{sec:3Q}). 
In Sec.~\ref{sec:energy}, we discuss 
the origin of the directional preference of DWs from 
 the wave-number dependence of 
 the bare susceptibility $\chi^0_{\bf q}$. 
Section~\ref{sec:conclusion} is devoted to summary.

\section{Model and method}
\label{sec:Model and method}
In this section, we present the model and method that we use. 
After introducing the Kondo lattice model 
in Sec.~\ref{sec:KLM}, 
we discuss the effective interaction between the localized spins mediated by mobile charges
in Sec.~\ref{sec:RKKY}. 
In Sec.~\ref{sec:KPM-LD}, we briefly review the KPM-LD method 
and introduce its extension by applying the SLL variant of LD.
We also review the method to evaluate physical quantities by the KPM in Sec.~\ref{sec:localKPM}. 

\subsection{Kondo Lattice Model}
\label{sec:KLM}

To investigate the DW problem in itinerant magnets,
we consider a minimal model describing the interplay between spin and charge degrees of freedom, 
the Kondo lattice model with classical localized spins. 
The Hamiltonian is given by 
\begin{align}
\label{eq:KLM}
\hat{\mathcal{H}} = -t \sum_{\langle i, j\rangle, \sigma} (\hat{c}^{\dagger}_{i\sigma}\hat{c}^{\;}_{j \sigma}+ {\rm H.c.})-J \sum_i \hat{\mathbf{s}}_i \cdot \mathbf{S}_i - D\sum_i(S^z_i)^2, 
\end{align}
where $\hat{c}^{\dagger}_{i\sigma}$ ($\hat{c}^{\;}_{i \sigma}$) is a creation (annihilation) operator of a mobile charge 
at site $i$ and spin $\sigma$, 
$\hat{\bf s}_i = \frac12 \sum_{\sigma \sigma'} \hat{c}^{\dagger}_{i\sigma} \bm{\sigma}_{\sigma \sigma'} \hat{c}^{\;}_{i \sigma'}$ 
is the spin operator of a mobile charge [$\bm{\sigma}=(\sigma^x,\sigma^y,\sigma^z)$ is the vector of Pauli matrices], ${\bf S}_i$ denotes a classical localized spin at site $i$ 
whose amplitude is normalized as $|{\bf S}_i|=1$.  
In the following study, we consider the model on square and triangular lattices; 
the sum of $\langle i,j \rangle$ is taken over the nearest-neighbor 
sites on each lattice. 
The first term in Eq.~(\ref{eq:KLM}) represents the hopping of mobile charges between the nearest-neighbor sites 
with the amplitude $-t$. 
The second term shows the onsite exchange coupling between the localized spins and spin degree of freedom of mobile charges 
with the coupling constant $J$ 
(the sign of $J$ is irrelevant for classical localized spins). 
The third term represents a uniaxial spin anisotropy with the amplitude $D$,  
which comes from, e.g., the relativistic spin-orbit coupling. 
Hereafter, we take $t=1$ and $a=1$ (lattice constant) as energy and length units, respectively.

In the Kondo lattice model in Eq.~(\ref{eq:KLM}), 
the coupling to mobile charges leads to an effective magnetic interaction between localized spins.
In the strong coupling case ($J\gg t$), $\langle\hat{{\bf s}}_i\rangle$  and ${\bf S}_i$ are almost 
in parallel, and
the effective hopping amplitude of mobile charges depends on the relative angle of neighboring localized spins\cite{PhysRev.82.403,PhysRev.100.675}. 
As a consequence,
the ferromagnetic ordering is favored to maximize the kinetic energy of mobile charges.
The effective ferromagnetic interaction between localized moments is called the double-exchange interaction\cite{PhysRev.82.403}.
On the other hand, when $J \ll t$, the effective magnetic interaction becomes complicated with oscillating sign depending on the distance between the localized spins. 
The weak coupling case will be discussed in the following sections.

\subsection{RKKY Interaction}
\label{sec:RKKY}

Hereafter, 
we investigate  magnetic DWs in the weak $J$ region of the Kondo lattice model, where various magnetic orderings are expected to occur owing to the effective magnetic interaction mediated by mobile charges as discussed below.
In the weak coupling limit ($J\ll t$), the effective magnetic interaction 
is derived by the second-order perturbation in terms of $J$, which is written as
\begin{equation}\label{eq:RKKY}
\mathcal{H}^{\rm RKKY} = - 
\frac{J^2}{4} 
\sum_{\bf q}\chi^0_{\bf q}|{\bf S}_{\bf q}|^2.
\end{equation}
Here, ${\bf S}_{\bf q}$ is the Fourier transform of ${\bf S}_i$ 
given by 
\begin{equation}
 {\bf S}_{\bf q} = \frac{1}{\sqrt{N}}\sum_j  {\bf S}_j e^{i{\bf q}\cdot{\bf r}_j}, 
\end{equation}
where $N$ is the number of sites and ${\bf r}_j =(r_j^x,r_j^y)
$ is the position of the site $j$.
In Eq.~(\ref{eq:RKKY}), $\chi^0_{\bf q}$ is the bare magnetic susceptibility of mobile charges,
\begin{equation}\label{eq:susceptibility}
\chi^0_{\bf q}=-\frac{1}{N}\sum_{\bf k}\frac{f(\varepsilon_{{\bf k}+{\bf q}}) - f(\varepsilon_{\bf k})}{\varepsilon_{{\bf k}+{\bf q}} - \varepsilon_{\bf k}},
\end{equation}
where $f(\varepsilon)$ is the Fermi distribution function
and $\varepsilon_{\bf k}$ is the dispersion relation of mobile charges with a  wave number ${\bf k} 
=(k_x,k_y)$
given by
\begin{equation}
\varepsilon_{\bf k} = -2\left(\cos k_x + \cos k_y\right) 
\end{equation}
for the square lattice and
\begin{align}
\varepsilon_{\bf k} = -2\Big[\cos k_x  &+ \cos\big(-\frac{k_x}{2} + \frac{\sqrt{3}k_y}{2}\big) \nonumber \\
& + \cos\big(-\frac{k_x}{2} -\frac{\sqrt{3}k_y}{2}\big)\Big] 
\end{align}
for the triangular lattice. 

The interaction in Eq.~(\ref{eq:RKKY}) is called the Ruderman-Kittel-Kasuya-Yosida (RKKY) 
interaction\cite{PhysRev.96.99,Kasuya01071956,PhysRev.106.893}. 
The wave number dependence gives rise to a long-range interaction in real space  
with oscillating sign.
Equation~(\ref{eq:RKKY}) provides us a good insight into the magnetic ground state in the Kondo lattice model:
as $|{\bf S}_{\bf q}|^2 \geq 0$ 
and $\sum_{\bf q}|{\bf S}_{\bf q}|^2=N$, 
the ground state will develop a magnetic order specified by the wave number(s) for which 
$\chi^0_{\bf q}$ is maximized.

\subsection{Modified KPM-LD Method}
\label{sec:KPM-LD}
To study the magnetic and electronic states in the Kondo lattice model,
we employ an efficient numerical technique, the KPM-LD method\cite{PhysRevB.88.235101}. 
This technique is widely applicable to the systems in which noninteracting 
fermions are coupled to classical degrees of freedom,  
and enables us to perform an unrestricted simulation in large-size systems.
In the following, we briefly introduce the KPM-LD technique and its most recent modifications that we use here.

In the Kondo lattice model in Eq.~(\ref{eq:KLM}), the partition function $Z$ is given by 
\begin{align}
 Z &= {\rm Tr}_{\{{\bf S}_i\}}  {\rm Tr}_{\{\hat{c}^{\;}_i\}} \exp\{ - [\mathcal{H}(\{{\bf S}_i\}) - \mu\sum_{i\sigma}\hat{c}^\dagger_{i\sigma}\hat{c}^{\;}_{i\sigma}] 
 /T\}\label{eq:Z_01}\\
 &=  {\rm Tr}_{\{{\bf S}_i\}} \exp[-\Omega(\{{\bf S}_i\})/T],\label{eq:Z_02}
\end{align}
where $\mu$ is the chemical potential and $T$ is temperature (we set the Boltzmann constant $k_{\rm B}=1$). 
$\Omega(\{{\bf S}_i\})$ is the grand potential for a spin configuration $\{{\bf S}_i\}$, which is given by
\begin{align}
\Omega(\{{\bf S}_i\}) 			&= -\int d\varepsilon \rho(\varepsilon;\{{\bf S}_i\}) T\log\{1 + \exp[-(\varepsilon - \mu)/T]\}.
\end{align}
Here, $\rho(\varepsilon;\{{\bf S}_i\})$ is the density of states, 
\begin{align}\label{eq:dos}
\rho(\varepsilon;\{{\bf S}_i\})	&= \frac{1}{N}\sum_j \delta(\varepsilon - \varepsilon_j(\{{\bf S}_i\})),
\end{align}
where $\varepsilon_j(\{{\bf S}_i\})$ is the 
$j$th eigenvalue of the Hamiltonian in Eq.~(\ref{eq:KLM}) for the spin configuration $\{{\bf S}_i\}$.
In the KPM-LD algorithm, the trace over ${\{\hat{c}^{\;}_i\}}$ in Eq.~(\ref{eq:Z_01}) is evaluated by the KPM and
the trace over ${\{{\bf S}_i\}}$ is evaluated by the stochastic sampling of the LD.

In the KPM for the trace over $\{\hat{c}_i\}$, the density of states in Eq.~(\ref{eq:dos}) is expanded by the Chebyshev polynomials as 
\begin{equation}\label{eq:dos_KPM}
\rho(x;\{{\bf S}_i\}) \simeq \sum_{m=0}^{M} g_m^{\rm J}w_m(x)\mu_m T_m(x),
\end{equation}
where $T_m(x)$ is the $m$th-order Chebyshev polynomial. 
Here, the energy $\varepsilon$  
is scaled so that all the eigenvalues are within the range $x \in [-1, 1]$, 
$w_m(x)$ is the weight function for Chebyshev polynomials given by $w_m(x) = (2-\delta_{0,m})/(\pi\sqrt{1-x^2})$, 
and $\mu_m$ is the Chebyshev moment given by $\mu_m = {\rm Tr} [T_m(\mathcal{H}(\{{\bf S}_i\}))]$.
Following the previous studies, 
we take the summation in the trace over random vectors, 
instead of the complete basis set\cite{RevModPhys.78.275}. 
To improve the accuracy of the stochastic KPM approximation, we select a set of correlated random vectors using a matrix probing technique\cite{preparation}  
inspired by Ref.~\onlinecite{tang2012probing}. 
In Eq.~(\ref{eq:dos_KPM}), $M$ is the truncation number for the Chebyshev polynomial expansion, and $g_m^{\rm J}$ is the so-called Jackson kernel which suppresses the error due to the truncation\cite{RevModPhys.78.275}. 

Meanwhile, 
in the LD that generates  spin configurations for the trace over $\{{\bf S}_j\}$, 
we modified the algorithm from 
the original one in Ref.~\onlinecite{PhysRevB.88.235101} 
by applying the SLL equation \cite{0953-8984-20-31-315203}.
The SLL equation is 
\begin{equation}\label{eq:sLLG}
\frac{d {\bf S}_i}{d\tau} = -{\bf S}_i\times{\bf H}_i - \alpha{\bf S}_i\times\left({\bf S}_i\times{\bf H}_i\right).
\end{equation}
The first and second terms represent a precession and dumping of ${\bf S}_i$, respectively; 
${\bf H}_i$ is the effective magnetic field for ${\bf S}_i$  
given by
\begin{equation}\label{eq:sLLG2}
{\bf H}_i = -\frac{\partial \Omega(\{{\bf S}_j\})}{\partial {\bf S}_i} + {\bf h}_i(\tau, T).
\end{equation}
Here, the first term on the RHS is numerically evaluated 
by the automatic differentiation transformation~\cite{PhysRevB.88.235101}, and
${\bf h}_i$ represents thermal fluctuations, which satisfies 
$\langle h^{\nu_1}_i(\tau_1, T)h^{\nu_2}_j(\tau_2, T)\rangle_\tau = 2T\delta(\tau_1-\tau_2)\delta_{ij}\delta_{{\nu_1}{\nu_2}}$,
where $\langle \cdots\rangle_\tau$ is the time average and $h_i^\nu$ is the $\nu$ component of ${\bf h}_i$.

For the time evolution, we use the Heun 
integration scheme, a type of predictor-corrector method, to achieve
 second-order accuracy with respect to $\Delta \tau$ 
(time interval of the update of spins). 
Note that the SLL equation, 
Eq.~(\ref{eq:sLLG}), preserves the length of spins, i.e., $|{\bf S}_i|=1$. 
However, numerical errors in the Heun scheme violate this condition at order $\mathcal O(\Delta \tau^2)$. 
To improve accuracy, we rescale the spin lengths
at each time step, which implements the so-called Heun+projection scheme\cite{0953-8984-22-17-176001}.

The computational cost of the modified KPM-LD simulation for an update of all the spins is $\mathcal{O}(N)$, similar to the original KPM-LD~\cite{PhysRevB.88.235101}.
This is much smaller than the cost of the conventional Monte Carlo algorithm, $\mathcal{O}(N^4)$\cite{Yunoki_PhysRevLett.80.845}. 
This drastic reduction of computational cost enables the simulation of systems with $\sim 10^4$ sites, 
whereas the conventional algorithm would be limited to several hundred sites.  

In the following, we use the modified KPM-LD method introduced above. 
In the calculations, 
we set $\alpha=1$ and $\Delta \tau=20$ 
in the SLL equation\cite{dtau}, 
and perform the Chebyshev polynomial expansion up to 
 $M=2000$ with using $144$ correlated random vectors for the KPM.
We perform the simulation at $T$ = 0 starting from a random spin configuration, 
which corresponds to a sudden quench from the high-$T$ limit to zero $T$. 
In the following, we show the results for the system size $N=120^2$ for both the square- and triangular-lattice cases. 
We confirmed that qualitatively the same results are obtained for 
several different random samples and 
for smaller size systems, $N=60^2$ and $90^2$. 
In the simulation, we use General-Purpose computing on Graphics Processing Units (GPGPU) 
to perform the sparse matrix operations required for the KPM approximation of the effective field, Eq. (\ref{eq:sLLG2}).

\subsection{Physical Quantities}
\label{sec:localKPM}
In this section, we show how to evaluate physical quantities for mobile electrons by the KPM\cite{PhysRevLett.73.1039,RevModPhys.78.275},
such as local charge and current densities. 
The expectation value of an operator $\hat A$ 
for a given spin configuration $\{{\bf S}_j\}$  
is obtained as
\begin{align}
 \langle \hat A \rangle &= \sum_i \langle i| \hat A | i\rangle f(\varepsilon_i(\{{\bf S}_j\})) = \int d\varepsilon A(\varepsilon;\{{\bf S}_j\}) f(\varepsilon),
\end{align}
where $\{ |i\rangle\}$ is the complete set of single-particle electron eigenstates of the Kondo lattice model for $\{{\bf S}_j\}$, and $A(\varepsilon; \{{\bf S}_j\})$ is defined by
\begin{align}
A(\varepsilon; \{{\bf S}_j\}) &= \sum_i \delta(\varepsilon- \varepsilon_i(\{{\bf S}_j\})) \langle i| \hat A | i\rangle\nonumber\\
 &= \sum_i \langle i|\delta(\varepsilon- \hat{\mathcal{H}}(\{{\bf S}_j\}))  \hat A | i\rangle.
\end{align}
Similar to Eq.~(\ref{eq:dos_KPM}), we can estimate $A(\varepsilon; \{{\bf S}_j\})$ by the 
KPM as
\begin{align}
A(x; \{{\bf S}_j\}) \simeq \sum_{m=0}^M  g_m^{\rm J} w_m(x)\mu^A_mT_m(x),
\label{eq:A_expansion}
\end{align}
where $\mu^A_m$ is the moment in the Chebyshev polynomial expansion. 
Using the orthonormality
of $T_m(x)$, $\mu^A_m$ is obtained as 
\begin{align}
\mu^A_m &= \int dx T_m(x) A(x; \{{\bf S}_j\})\nonumber\\
&=\int dx \sum_i \langle i| T_m(x)\delta(x - \hat{\mathcal{H}}(\{{\bf S}_j\})) \hat{A}|i\rangle\nonumber\\
&=\sum_i \langle i| T_m(\hat{\mathcal{H}}(\{{\bf S}_j\}))\hat{A}|i\rangle \nonumber \\
&= {\rm Tr}[T_m(\hat{\mathcal{H}}(\{{\bf S}_j\}))\hat{A}].\label{eq:loc_KPM}
\end{align}

Equation~(\ref{eq:loc_KPM}) is calculated by massive vector-matrix products, 
whose computational cost is $\mathcal{O}(N^2)$ when the complete basis set is used. 
However, if $\hat{A}$ is a local operator, 
e.g., the one defined for a site, bond, or plaquette, 
 it only requires $\mathcal{O}(N)$ cost 
because the sum of $i$ in Eq.~(\ref{eq:loc_KPM})  
is limited to the sites where $\hat{A}$ is defined.

\section{NUMERICAL RESULTS}
\label{sec:results}
In this section, we examine the shape of magnetic domains in the Kondo lattice model 
 by the modified KPM-LD simulation.
In Secs.~\ref{sec:AFM} and \ref{sec:120}, 
we present the results for the collinear N\'eel state 
 on the square lattice and the coplanar $120^{\circ}$ state 
 on the triangular lattice, respectively.
In Sec.~\ref{sec:3Q}, we show the results for noncoplanar triple-$Q$ states with scalar chiral ordering
which are stabilized near 1/4 and 3/4 fillings on the triangular lattice.
We also present how the spontaneous electric current by the spin Berry phase mechanism flows along the DWs in each case. 
In each section, we present the wave-number dependence of the bare susceptibility $\chi^0_{\bf q}$ in Eq.~(\ref{eq:susceptibility}). The relation between $\chi^0_{\bf q}$ and  
the anisotropy of DWs in real space will be discussed in Sec.~\ref{sec:energy}.

\subsection{Collinear N\'eel State}
\label{sec:AFM}
\begin{figure}[h]
\begin{center}
\includegraphics[width=1.0 \hsize]{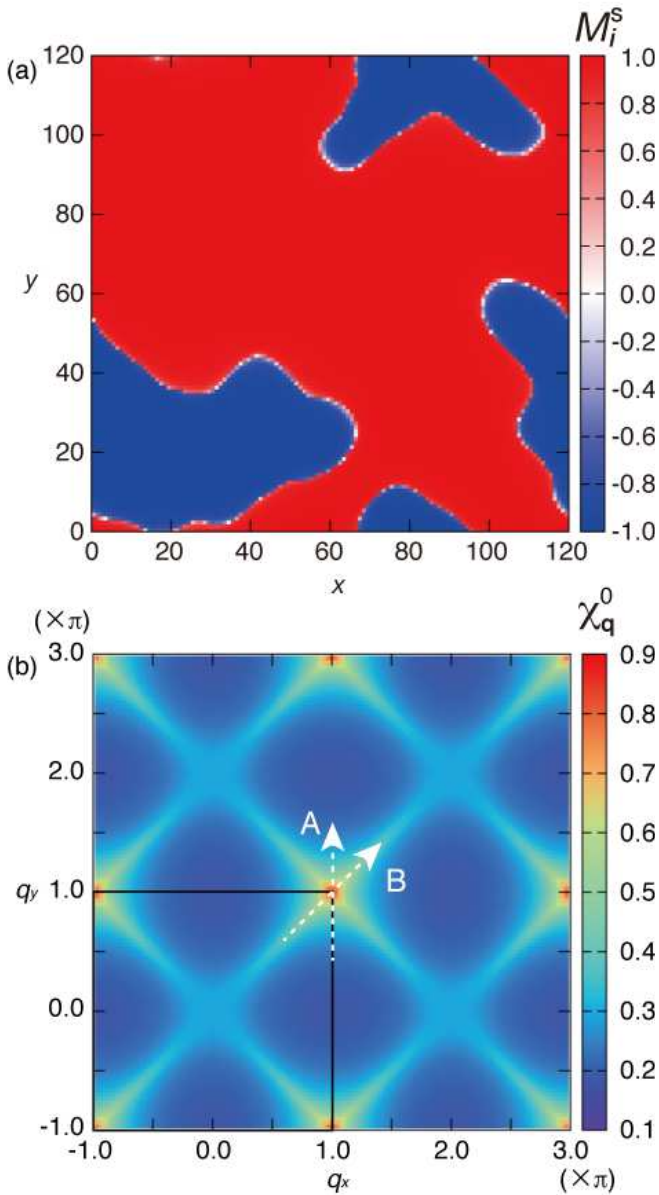} 
\caption{(Color online) (a) Real-space distribution of the staggered magnetization $M^{\rm s}_i$ obtained by the modified KPM-LD simulation 
for the Kondo lattice model on the square lattice at $\mu=0$ ($n\sim 0.5$), $J=0.2$, $D=0.005$, and $N=120^2$. 
(b) Bare susceptibility $\chi^0_{\bf{q}}$ at $\mu=0$ in the momentum space. 
The black square represents the 1st BZ. 
The arrows A and B denote the cuts along which we evaluate 
the second derivative of $\chi^0_{\bf q}$; see Sec.~\ref{sec:energy} for details.
\label{fig:Neel}}
\end{center}
\end{figure}

Let us first discuss the case of
the N\'eel state, which is a collinear antiferromagnetic (AFM) state. 
It is stabilized in the Kondo lattice model on the square lattice near half filling because of the perfect nesting of the Fermi surface with the $(\pi, \pi)$ ordering vector. 
The spatial pattern of the localized spins is represented by 
\begin{align}
{\bf S}_i = \left(
\begin{array}{c} 
0 \\ 
0 \\ 
(-1)^{r_i^x + r_i^y} 
\\ 
\end{array}
\right). 
\end{align}
We investigate what type of the domain structures is realized in the N\'eel state by the modified KPM-LD simulation. 

Figure~\ref{fig:Neel}(a) shows a snapshot of 
domain structure in the N\'eel state
 obtained by the modified KPM-LD simulation  
for the square lattice system with $N=120^2$ under periodic boundary conditions. 
We set $\mu=0$ to realize the half filling, $n\sim 0.5$ [$n = \sum_{i\sigma}\langle\hat{c}^\dagger_{i\sigma}\hat{c}^{\;}_{i\sigma}\rangle/(2N)$], and take $J=0.2$. 
Figure~\ref{fig:Neel}(a) represents the real-space configuration of the staggered magnetization, $M^{\rm s}_i=(-1)^{i_x + i_y}S_i^z$, at $\tau=2\times10^3$; 
we here introduced a
small positive $D=0.005$ to observe the domains in a clear form. 
In the figure, there are three domains 
separated by the DWs. 
The DWs have overall round shapes, while we can see a weak preference along the 
diagonal directions compared to the horizontal and vertical directions.

Figure~\ref{fig:Neel}(b) shows the corresponding bare susceptibility $\chi^0_{\bf q}$ defined in Eq.~(\ref{eq:susceptibility}), 
calculated at $\mu=0$ and $T=0.05$ for $N=600^2$. 
As expected from the perfect nesting property, $\chi^0_{\bf q}$ exhibits a sharp peak 
at ${\bf q}=(\pi, \pi)$ in the first Brillouin zone (1st BZ), which grows to a $\delta$ function in the limit of $T\to 0$ and $N\to \infty$. 
The peak is weakly anisotropic 
in the momentum space between the vertical (A) and diagonal (B) directions.
The relation between the structure of $\chi^0_{\bf q}$ near the peaks and 
the shape of DWs will be discussed in Sec.~\ref{sec:energy}.

\subsection{Coplanar 120$^\circ$ State}
\label{sec:120}

\begin{figure}[h]
\begin{center}
\includegraphics[width=1.0 \hsize]{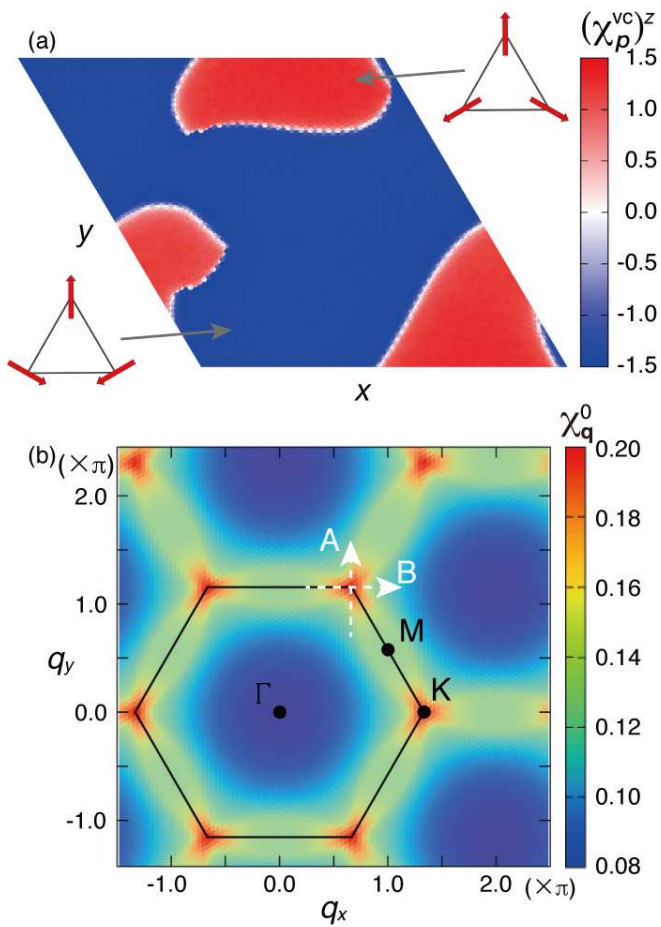} 
\caption{(Color online) (a) Real-space distribution of  the $z$ component of the vector chirality ${\bm \chi}^{\rm vc}_p$ obtained by the modified KPM-LD simulation on the triangular lattice at $\mu=-0.95$ ($n\sim0.31$), $J=0.2$, and $D=-0.005$ for $N=120^2$. 
The inset shows the schematic pictures of spin patterns for the positive and negative $({\chi}^{\rm vc}_p)^z$ on triangular plaquettes.
 (b) Bare susceptibility $\chi^0_{\bf{q}}$ at $\mu=-0.95$ in the momentum space.
 The black hexagon represents the 1st BZ.  
 The arrows A and B denotes the cuts along which 
  we evaluate the second derivative of $\chi^0_{\bf q}$; see Sec.~\ref{sec:energy} for details.
\label{fig:120}}
\end{center}
\end{figure}

Next, we consider the case for a noncollinear but coplanar
120$^\circ$ AFM state on the triangular lattice. 
The spin configuration is typically described by 
\begin{align}
{\bf S}_i = \left(
\begin{array}{c} 
\cos({\bf Q}_{\rm K}\cdot{\bf r}_i) \\ 
\sin({\bf Q}_{\rm K}\cdot{\bf r}_i)  \\ 
0\\ 
\end{array}
\right),
\end{align}
where ${\bf Q}_{\rm K}$  
denotes the wave number at the K point in the 1st BZ: ${\bf Q}_{\rm K} = (\frac 43 \pi, 0)$  
or $(\frac 23\pi, \frac{2}{\sqrt{3}}\pi)$ [see Fig.~\ref{fig:120}(b)]. 
The 120$^\circ$ AFM order is regarded as a ferroic order of the vector chirality defined as 
\begin{align}
{\bm \chi}^{\rm vc}_p= {\bf S}_{p_1}\times{\bf S}_{p_2} + {\bf S}_{p_2}\times{\bf S}_{p_3}+ {\bf S}_{p_3}\times{\bf S}_{p_1}, 
\label{eq:vc}
\end{align}
where $p_1$, $p_2$, and $p_3$ are the sites on each triangular plaquette $p$ in a counterclockwise direction. 
We use the vector chirality to characterize the domains in the 120$^\circ$ AFM state.

Figure~\ref{fig:120}(a) shows a snapshot of the vector chirality
obtained by the modified KPM-LD simulation for the triangular lattice system with $N=120^2$ under periodic boundary conditions.
We set $\mu=-0.95$ ($n\sim 0.31$) to stabilize the 120$^\circ$ AFM order, and take $J=0.2$.
The plot shows the $z$ component of the vector chirality in Eq.~(\ref{eq:vc}), $(\chi_p^{\rm vc})^z$,  
at $\tau=3.2\times10^4$; we introduced a small negative $D=-0.005$ to clearly observe the domains.
There are two domains with positive and negative $(\chi^{{\rm vc}}_p)^z$ 
separated by one DW, as shown in Fig.~\ref{fig:120}(a).
The typical spin patterns in each domain are shown in the inset.
The DW has a round shape and does not show the strong preference in the direction.

Figure~\ref{fig:120}(b) shows the corresponding bare susceptibility $\chi^0_{\bf q}$ in Eq.~(\ref{eq:susceptibility})  
evaluated at $\mu=-0.95$ and $T=0.05$ for $N=600^2$.  
$\chi^0_{\bf q}$ exhibits a distinct peak at the K point in the 1st BZ, 
which is consistent with the stabilization of $120^{\circ}$ ordering. 
The structure of $\chi^0_{\bf q}$ near the peak is nearly isotropic with threefold rotational symmetry.
We will return to this point in Sec.~\ref{sec:energy}.

\subsection{Noncoplanar Triple-$Q$ States}
\label{sec:3Q}

\begin{figure*}[h]
\begin{center}
\includegraphics[width=1.0\hsize]{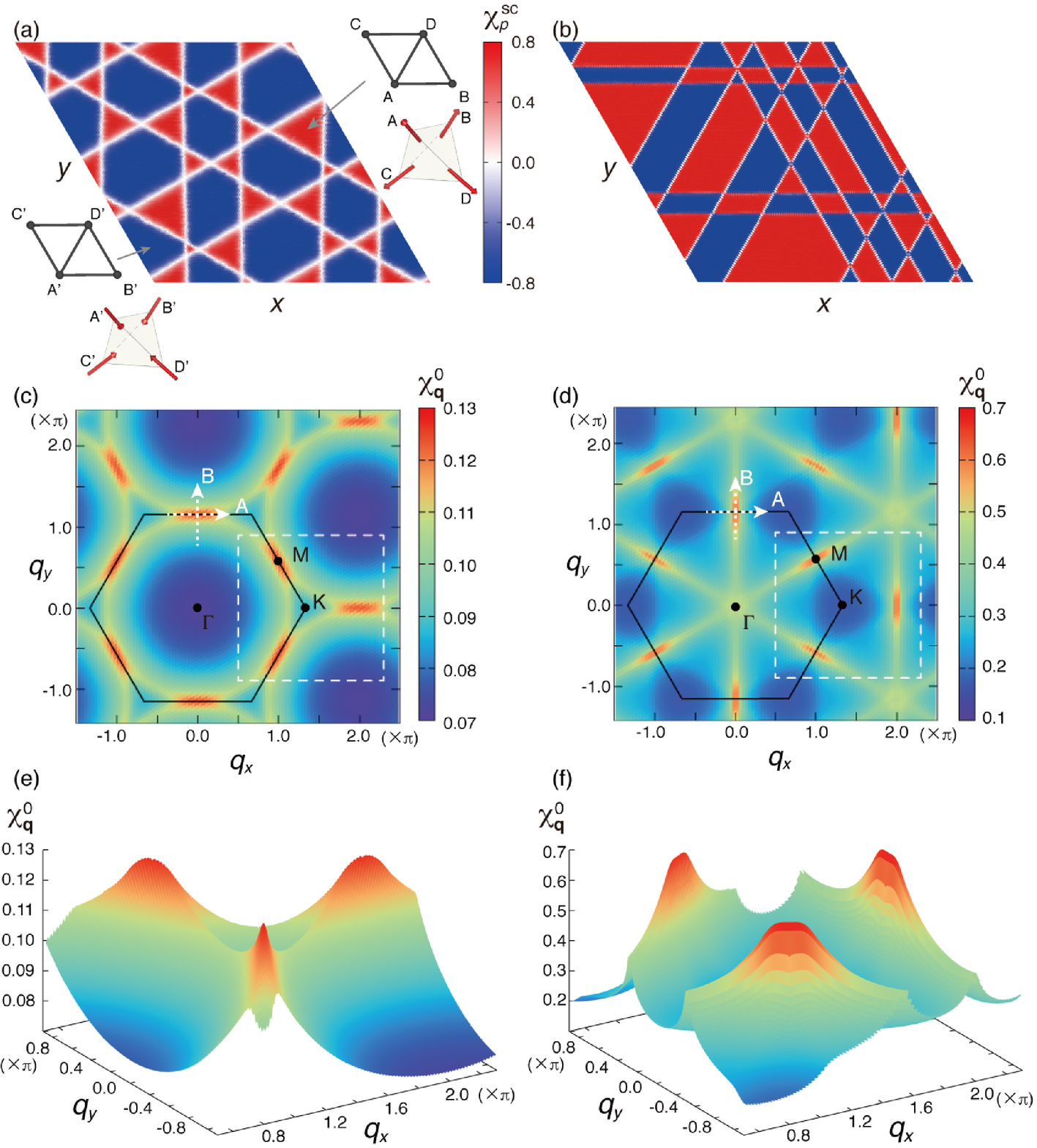} 
\caption{(Color online) Real-space distributions of spin scalar chirality $\chi^{ \rm sc}_p$ obtained by the modified KPM-LD simulation on the 
triangular lattice for $J=0.2$, $D=0$, $N=120^2$, and (a) $\mu=-2$ ($n\sim0.22$) and (b) $\mu=2$ ($n\sim 0.75$). 
The inset in (a) shows the schematic pictures of spin patterns in the noncoplanar triple-$Q$ states, corresponding to the positive and negative ${\chi}^{\rm sc}_p$ domains. The arrows at the corners of the tetrahedra represent the spin directions at the corresponding sites. 
Bare susceptibilities $\chi^0_{\bf q}$  for (c) $\mu=-2$ and (d) $\mu=2$.
The black hexagon represents the 1st BZ in each figure.
The arrows A and B show the cuts along  which we evaluate the second derivative of $\chi^0_{\bf q}$; 
see Sec.~\ref{sec:energy} for details.
(e), (f) Three-dimensional plots of  $\chi^0_{\bf q}$ in the dashed squares in (c) and (d), respectively. 
\label{fig:3Q}
}
\end{center}
\end{figure*}

In this section, we consider the case for a triple-$Q$ AFM state.
The triple-$Q$ state is characterized by three ordering vectors,
whose spin pattern is typically given by
\begin{align}
{\bf S}_i = \frac{1}{\sqrt{3}}\left(
\begin{array}{c} 
\cos({\bf Q}_{{\rm M}_1}\cdot{\bf r}_i) \\ 
\cos({\bf Q}_{{\rm M}_2}\cdot{\bf r}_i)  \\ 
\cos({\bf Q}_{{\rm M}_3}\cdot{\bf r}_i) \\ 
\end{array}
\right).
\end{align}
Here, ${\bf Q}_{{\rm M}_\ell}$ ($\ell=1, 2, 3$) denote the three ordering vectors: ${\bf Q}_{{\rm M}_1} = (\pi, \frac{1}{\sqrt{3}}\pi)$, ${\bf Q}_{{\rm M}_2} = (0, \frac{2}{\sqrt{3}}\pi)$, and ${\bf Q}_{{\rm M}_3}=(-\pi, \frac{1}{\sqrt{3}}\pi)$, which correspond to the M points in the 1st BZ.
Thus, the triple-$Q$ order has a noncoplanar spin pattern with four-sublattice unit cell, as shown in the inset of Fig.~\ref{fig:3Q}(a). 
This peculiar order takes place in the Kondo lattice model on the triangular lattice near $3/4$ and $1/4$ fillings:
the former is stabilized by the perfect nesting of the Fermi surface at $3/4$ filling
\cite{PhysRevLett.101.156402},  
while the latter by a partial nesting near $1/4$ filling
\cite{PhysRevLett.108.096401, PhysRevB.90.060402}.   
The noncoplanar triple-$Q$ states 
accompany a ferroic order of scalar chirality, 
\begin{align}
\chi^{\rm sc}_p = {\bf S}_{p_1}\times{\bf S}_{p_2} \cdot {\bf S}_{p_3},
\end{align}
defined on the 
triangular plaquette $p$, as in Eq.~(\ref{eq:vc}).
$\chi^{\rm sc}_p$ acts as a $Z_2$ variable even when the Hamiltonian preserves the ${\rm SO}(3)$ symmetry. 
We use $\chi^{\rm sc}_p$ to characterize the magnetic domains in the triple-$Q$ AFM states near $1/4$ and $3/4$ fillings in the following.

Figure~\ref{fig:3Q}(a) shows a snapshot of the scalar chirality in the triple-$Q$ ordered state near $1/4$ filling,
obtained by the modified KPM-LD simulation for the triangular lattice system with $N=120^2$
under periodic boundary conditions.
We set $\mu=-2$ ($n\sim 0.22$) and take $J=0.2$ and $D=0$. 
The figure represents the real-space configuration at $\tau=8\times10^4$. 
Unlike the cases of the N\'eel and $120^{\circ}$ states, 
DWs have a strong preference in their directions; they prefer three directions, all of which are  
 perpendicular to the nearest-neighbor bonds of the triangular lattice. 

Figure~\ref{fig:3Q}(b) shows 
the result for the triple-$Q$ state near $3/4$ filling.
We set $\mu=2$ $(n\sim 0.75)$ 
and take the same values for the other parameters as in Fig.~\ref{fig:3Q}(a).
While the system exhibits several domains 
as in Fig.~\ref{fig:3Q}(a), the directions of DWs exhibit a different preference from the previous case: they prefer the three directions parallel to the bonds. 

Figures~\ref{fig:3Q}(c) and \ref{fig:3Q}(d) show the bare susceptibilities $\chi^0_{\bf q}$ 
corresponding to the $1/4$- and $3/4$-filling cases, respectively. 
We take
$\mu=-2$ and $T=0.05$ in Fig.~\ref{fig:3Q}(c), and $\mu=2$ and $T=0.01$ in Fig.~\ref{fig:3Q}(d); $N=600^2$ in both cases. 
For both fillings, $\chi^0_{\bf q}$ 
 exhibits distinct peaks at the M points in the 1st BZ, consistent with the emergence of the triple-$Q$ order. 
However, the 
wave number dependences  
around the peaks are totally different from each other. 
In the case of $1/4$ filling, the peak of $\chi^0_{\bf q}$ is broader along the direction A
compared to the perpendicular direction B, as shown in Fig.~\ref{fig:3Q}(c).
In contrast, in the case of $3/4$ filling, the peak of $\chi^0_{\bf q}$  
is much sharper along the direction A
than B, as shown in Fig.~\ref{fig:3Q}(d). 
To show the anisotropic structures of $\chi^0_{\bf q}$ near the peaks more clearly, 
we present the enlarged figures in the three-dimensional style
in Figs.~\ref{fig:3Q}(e) and \ref{fig:3Q}(f), corresponding to the white dashed areas in Figs.~\ref{fig:3Q}(c) and \ref{fig:3Q}(d), respectively.
We will discuss the relation between the wave number dependences of $\chi^0_{\bf q}$ and the distinct preferences in the direction of DWs in Sec.~\ref{sec:energy}.

\begin{figure*}[h]
\begin{center}
\includegraphics[width=1.0\hsize]{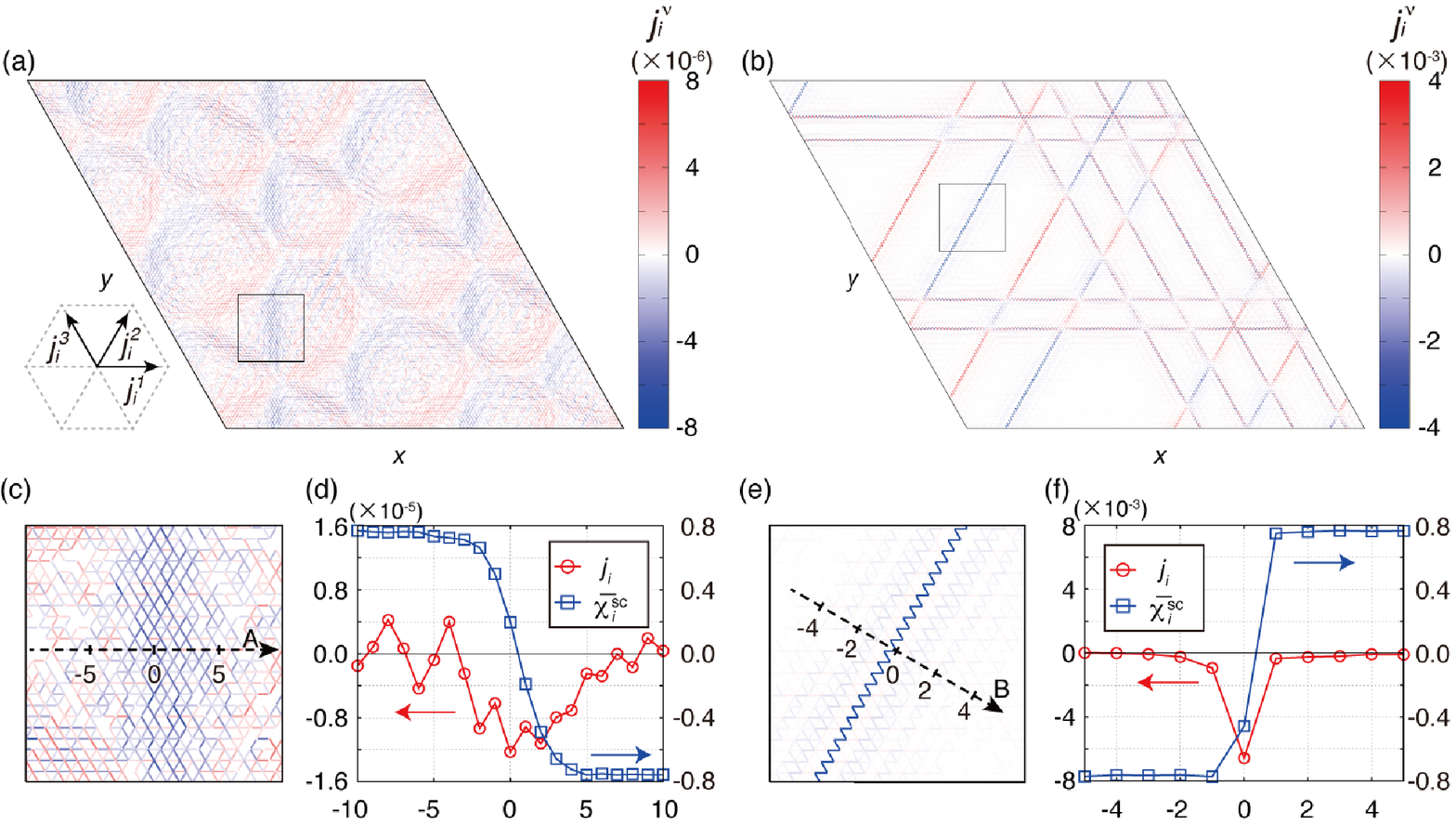} 
\caption{(Color online) (a), (b) Real-space distributions of the local current density ${j}_i^\nu$ on each bond 
for the spin configurations in Figs.~\ref{fig:3Q}(a) and \ref{fig:3Q}(b), respectively. 
(c), (e) Enlarged pictures of the square areas in (a) and (b), respectively.
(d), (f) Current density $j_i$ and spin scalar chirality $\bar{\chi^{\rm sc}_i}$ 
plotted along the arrows A and B 
in (c) and (e), respectively; see the text for details. 
\label{fig:current}
}
\end{center}
\end{figure*}

Besides the direction of the DWs, we 
note that the width of DWs is different between the two cases [see also Figs.~\ref{fig:current}(d) and \ref{fig:current}(f)].
The difference presumably stems from the different electronic structures. 
The triple-$Q$ state is metallic for the parameters used in Fig.~\ref{fig:3Q}(a), whereas it is insulating for Fig.~\ref{fig:3Q}(b). 
DWs might be thicker in the metallic system than the insulating one, reflecting 
the difference of the correlation length of electrons through the spin-charge coupling.

The difference is more directly observed in the electronic properties.
As the triple-$Q$ AFM states are ferroic ordered states of the spin scalar chirality, they exhibit spontaneous electric currents at the edges of the system
due to the spin Berry phase mechanism\cite{doi:10.7566/JPSJ.83.073706}. 
Such edge currents also appear at the DWs.
Figures~\ref{fig:current}(a) and \ref{fig:current}(b) show the  
real-space distributions of the local  current density corresponding to the spin states in Figs.~\ref{fig:3Q}(a) and \ref{fig:3Q}(b), 
respectively, obtained by the method in Sec.~\ref{sec:localKPM}.
Here, the local current density is defined as
\begin{align}
 j_{k}^\nu = \frac{1}{2i}\langle\hat{c}^\dagger_k\hat{c}^{\;}_l - \hat{c}^\dagger_l\hat{c}^{\;}_{k} \rangle,
\end{align}
where $k$ and $l$ are the nearest-neighbor sites along the $\nu$ direction ($\nu=1,2,3$), as shown in the inset of Fig.~\ref{fig:current}(a).
The color of each bond 
represents $j_{i}^\nu$ defined on the bond. 
As shown in Fig.~\ref{fig:current}(a) for the case of 1/4 filling, 
the electric current flows in a (counter)clockwise direction in the domains with positive (negative) $\chi^{\rm sc}_p$.
On the other hand, as shown in Fig.~\ref{fig:current}(b), the current flows in an opposite direction in the case of 3/4 filling. 
The directional difference is explained by the Berry curvature 
of the occupied bands at each filling\cite{PhysRevLett.101.156402,JPSJ.79.083711}. 
More interestingly, the two cases show a large difference in the spatial distribution of the current density.
In the case of 1/4 filling, 
the current density is distributed in a wide region 
in each domain, 
but in the case of 3/4 filling,  
it is spatially limited to the vicinity of the DWs.
This is presumably because the influence of DWs spreads over a wider range in the 1/4-filling metallic state than the 3/4-filling insulating one as the electron correlation length is longer in the former metallic case than the latter insulating case, as mentioned above for the thickness of the DWs. 

Let us look closer how the current density and spin scalar chirality change near DWs. 
Figure~\ref{fig:current}(c) shows the enlarged figure of Fig.~\ref{fig:current}(a) in the square region, where a vertical DW runs in the vertical direction in the center of the square [see Fig.~\ref{fig:3Q}(a)]. 
We plot the projected current density and spin scalar chirality along the arrow A in Fig.~\ref{fig:current}(c).
Here, we define the projected current density at site $i$, $j_i$, by the sum of  $j_k^\nu$ projected onto the upward direction along the DW [perpendicular to A in Fig.~\ref{fig:current}(c)]; we take the sum of the projected $j_k^\nu$ over six bonds connected to the site $i$. 
We also define the averaged scalar chirality at site $i$, $\bar{\chi_i^{\rm sc}}$, 
as the average of the spin scalar chirality on the six plaquettes including the site $i$.
Figure~\ref{fig:current}(d) shows the profiles of $j_i$ and $\bar{\chi_i^{\rm sc}}$ along the cut A in Fig.~\ref{fig:current}(c).
The result indicates that $\bar{\chi_i^{\rm sc}}$ changes smoothly from positive to negative across the DW, and 
$j_i$ exhibits a broad negative peak where $\bar{\chi_i^{\rm sc}}$ is suppressed.  
$j_i$ has a nonzero value over about ten sites 
around the DW, while it shows bumpy behavior presumably due to statistical fluctuations of the small quantity. 
The corresponding plots for the 3/4-filling case are shown in Figs.~\ref{fig:current}(e) and \ref{fig:current}(f).
In this case, $\bar{\chi_i^{\rm sc}}$  
changes rather sharply near the DW, 
and correspondingly, 
the negative peak of $j_i$ is much sharper than that in the 1/4-filling case; 
$j_i$  decays to zero much quicker than that in Fig.~\ref{fig:current}(c).  
We also note that the absolute value of $j_i$ at the DW is two orders of magnitude larger than that in Fig.~\ref{fig:current}(d).  

In the case of the triple-$Q$ state at $3/4$ filling, 
we note that the current near the DWs flows in the zigzag way,
 as shown in the enlarged figure 
 in Fig.~\ref{fig:current}(e); 
 the local currents are relatively small on the bonds along the DW direction, while they have substantial values along the other two directions near the DWs. 
This is because the localized spins are almost antiparallel 
on the bonds along the DWs, 
which suppresses the kinetic motion of mobile charges along this direction.

\section{Directional Preference of Domain Walls}
\label{sec:energy}

In the previous section, we found that the magnetic DWs may show directional preferences in some cases. 
In this section, we discuss the relationship between the structure of the bare susceptibility $\chi^0_{\bf q}$ 
and the directional preference of DWs.

Suppose the system 
shows a peak in $\chi^0_{\bf q}$ at $\bf q = {\bf q}^*$ and
a helical ordered state with the ordering wave vector ${\bf q}^*$  is realized in the ground state.
The following argument is straightforwardly generalized to the cases 
with multiple wave vectors, e.g., the triple-$Q$ states. 
In the helical ordered state, 
a typical spin pattern with a single DW 
can be described by a superposition of two helices, 
${\bf q}^*+d{\bf q}$ and ${\bf q}^*-d{\bf q}$, with equal weights, where  
$d{\bf q}$ described a small deviation from ${\bf q}^*$. 
This DW state exhibits two pairs of Bragg peaks in the spin structure factor as
\begin{align} 
|{\bf S}_{\bf q}| = 
\left\{ \begin{array}{ll}
    \sqrt{N}/2 & ({\bf q} = \pm{\bf q}^*\pm d{\bf q}) \\
    0 & {\rm (otherwise)}.
  \end{array} \right.
\end{align}
Note that ${\bf S}_{\bf q}$ satisfies the sum rule $\sum_{\bf q}|{\bf S}_{\bf q}|^2=N$.
The free energy of the RKKY interaction for the DW state is evaluated as 
\begin{align}
{\mathcal F}_{\rm DW} 
&= - \frac{J^2}{2} 
\left(\chi^0_{{\bf q}^*+d{\bf q}} {\bf S}_{{\bf q}^*+d{\bf q}}\cdot{\bf S}_{-{\bf q}^*-{d{\bf q}}}\right. \nonumber\\
  & \ \ \ \ \ \ \ \ \ \ \ \ \left. + \chi^0_{{\bf q}^*-{d{\bf q}}} {\bf S}_{{\bf q}^*-{d{\bf q}}}\cdot{\bf S}_{-{\bf q}^*+{d{\bf q}}}\right) \nonumber\\
&= -\frac{J^2N}{8}
\left(\chi^0_{{\bf q}^*+d{\bf q}}  + \chi^0_{{\bf q}^*-{d{\bf q}}} \right). 
\end{align}
Suppose this can be expanded by the small $d{\bf q}$, we obtain 
\begin{align}
{\mathcal F}_{\rm DW} = - \frac{J^2N}{4} 
\left(\chi^0_{{\bf q}^*}  +  \frac{1}{2}\frac{d^2 \chi^0_{\bf q}}{d{\bf q}^2}\bigg{|}_{{\bf q}={\bf q}^*}d{\bf q}^2 + \mathcal{O}(d{\bf q}^{3})\right).
\label{eq:energy_DW}
\end{align}
Note that the contribution linear to $d{\bf q}$  
vanishes because $d \chi^0_{\bf q}/d{\bf q}=0$ at ${\bf q}={\bf q}^*$. 
The first term in  Eq.~(\ref{eq:energy_DW}) corresponds to
the RKKY energy of the helical ordered state without the DW. 
Hence, the second term in Eq.~(\ref{eq:energy_DW}) describes the leading contribution from the DW.
Equation~(\ref{eq:energy_DW}) indicates that the creation of the DW always requires an energy cost 
as the second derivative of $\chi^0_{\bf q}$ with respect to ${\bf q}$ is always negative at ${\bf q}={\bf q}^*$ by definition 
as long as $\chi^0_{\bf q}$ is differentiable with respect to ${\bf q}$. 
The energy cost is minimized when we choose $d{\bf q}$ along the direction where 
$|d^2 \chi^0_{\bf q}/d{\bf q}^2|$ at ${\bf q}={\bf q}^*$ 
becomes smallest.
This suggests that the DW has a preference in the direction along which the peak of $\chi^0_{\bf q}$ decays most slowly.

The simple analysis well explains the numerical results obtained in Sec.~\ref{sec:results}.
For N\'eel and $120^{\circ}$ states in Secs.~\ref{sec:AFM} and \ref{sec:120}, 
$\chi^0_{\bf q}$ is rather isotropic in the momentum space, as shown in Figs.~\ref{fig:Neel}(b) and \ref{fig:120}(b). 
Indeed, the values of $d^2 \chi^0_{\bf q}/d{\bf q}^2$ in the directions
A and B are in the same order. 
This explains the fact that the DWs in these two states do not show a strong preference in their directions. 
As mentioned in Sec.~\ref{sec:AFM}, we noted that in the N\'eel case there is a small preference along the diagonal directions. 
This is also consistent with the fact that $d^2 \chi^0_{\bf q}/d{\bf q}^2$ in the B direction is smaller than that in the A direction, as shown in Fig.~\ref{fig:Neel}(b). 
On the other hand, for the triple-$Q$ states discussed in Sec.~\ref{sec:3Q}, $\chi^0_{\bf q}$ has strongly anisotropic structure, as shown in Figs.~\ref{fig:3Q}(c)-\ref{fig:3Q}(f). 
The anisotropy depends on the electron filling, which well explains the distinct preference of the DW directions found in the numerical simulations. 
In the 1/4-filling case,
$|d^2 \chi^0_{\bf q}/d{\bf q}^2|$ 
along the direction A is about ten times smaller than B.
This explains the reason why the DWs prefer the perpendicular directions to the nearest-neighbor bonds of the triangular lattice.
Meanwhile, in the 3/4-filling case, $|d^2 \chi^0_{\bf q}/d{\bf q}^2|$
along A is more than $100$ times larger than B, which is consistent with the formation of DWs along the bonds.

Let us make two remarks on the simple analysis in Eq.~(\ref{eq:energy_DW}).
The first one is on the expansion of $\chi^0_{\bf q}$ with respect to $d{\bf q}$. 
In some cases, 
$\chi^0_{\bf q}$ has a singular form with the $\delta$ functional peak at ${\bf q}={\bf q}^*$ in the zero-temperature limit. 
This occurs when the system is at the van Hove singularity with ${\bf q}={\bf q}^*$. 
In fact, this is the case for the N\'eel state 
and the 3/4-filling triple-$Q$ state.
Even in these cases, the simple analysis above may be applicable, as the singularity is smeared out at finite temperatures and the DWs obey the energetics in Eq.~(\ref{eq:energy_DW}) through the development of $\chi^0_{\bf q}$ while the annealing procedure.

The second point is on the higher-order contributions beyond RKKY. 
In the case of the triple-$Q$ states discussed in Sec.~\ref{sec:3Q}, 
we note that the contributions beyond the RKKY interaction 
plays an essential role 
in their stabilization mechanism\cite{PhysRevLett.108.096401}. 
Such beyond-RKKY contributions, however, are irrelevant in the directional preference of DWs, 
as the DW states with multiple-$Q$ ordering 
are also described by a superposition of the multiple-$Q$ states and the similar argument to the single-$Q$ helical state will apply to the states. 
On the other hand, the higher-order contributions will 
play a role in 
the spatial patterns of the spin texture near DWs.  
This needs more careful analysis, which is out of scope of the 
present study.

\section{Summary}
\label{sec:conclusion}

To summarize, we have investigated the formation of magnetic DWs  
through the spin-charge coupling by large-scale numerical simulation.  
We have studied the collinear N\'eel, coplanar 120$^\circ$, and noncoplanar triple-$Q$ states
in the Kondo lattice model with classical magnetic moments 
by the modified KPM-LD simulation at zero temperature 
starting from a random spin configuration, corresponding to the sudden quench. 
Although neither the N\'eel nor 120$^\circ$ state  
shows a strong preference in the direction of DWs, 
the triple-$Q$ states near 1/4 and 3/4 fillings 
exhibit distinct directional preferences depending on the filling.
In the 1/4-filling case, DWs run dominantly along the directions perpendicular to
the nearest-neighbor bonds of the triangular lattice, while in the 3/4-filling case, 
they strongly favor the directions parallel to the bonds, as observed  in the previous study\cite{PhysRevB.88.235101}.
We clarified that the directional preference of magnetic DWs in the weak-coupling region is predominantly determined by the electronic structure of mobile charges. 
This is rationalized by the fact that, in the weak-coupling region, the effective magnetic interaction, the so-called RKKY interaction, is given by the bare susceptibility $\chi^0_{\bf q}$, which is determined by the electronic band structure and the electron filling. 
While the ordering wave vector is determined by the peak of $\chi^0_{\bf q}$, we found that the directional preference of DWs is related with the wave-number dependence around the peak. 
When $\chi^0_{\bf q}$ is nearly isotropic 
 around the peak, DWs have overall round shapes. 
On the other hand, if $\chi^0_{\bf q}$ has distinct anisotropy, the directions along which the peak of  $\chi^0_{\bf q}$ decays most slowly correspond to the directions strongly preferred by DWs.
The former occurs in the collinear N\'eel and coplanar 120$^\circ$ states, and the latter in the two noncoplanar triple-$Q$ states. 
In particular, in the triple-$Q$ states, $\chi^0_{\bf q}$ shows the anisotropy in a different manner between the 1/4- and 3/4-filling states, which is reflected in the different preference of the DW directions.

In the noncoplanar triple-$Q$ states near 1/4 and 3/4 fillings, 
we have calculated the spontaneous currents induced along the DWs
 through the spin Berry phase mechanism. 
We confirmed that, reflecting the opposite sign of the Chern numbers, 
the current flows in the opposite directions at 1/4 and 3/4 fillings, 
as predicted in the previous studies\cite{PhysRevLett.101.156402,JPSJ.79.083711}. 
In addition, we found that 
the current density is spatially distributed 
in a wide region in the 1/4-filling case, 
while it is rather confined in the vicinity of DWs in the 3/4-filling case.
We observed the similar tendency in the width of the magnetic DWs (the region in which the triple-$Q$ order is disturbed);
the width of DWs is thicker in the 1/4-filling case than the 3/4-filling case.
The distinct behavior is presumably explained by the difference in the electronic states. 
The former is metallic, while the latter is insulating; the electron correlation length is longer in the former, which affects the electronic and magnetic properties in a wider region around the DWs.

Our analysis provides  
a simple mechanism of the directional preference of magnetic DWs in itinerant magnets. 
In this mechanism, the shapes of magnetic DWs are largely affected by   
the electronic band structure of mobile charges.
In other words, our mechanism is based on the momentum-space (itinerant) picture, which is different from that by 
the conventional real-space (localized) picture originating from 
the crystalline shape, magnetic anisotropy, and so on. 
As mentioned in Sec.~\ref{sec:Introduction}, recently, the interesting magnetic DWs  
were observed in several itinerant magnets showing metal-insulator transitions with peculiar magnetic ordering, such as Cd$_2$Os$_2$O$_7$\cite{PhysRevLett.114.147205} and Nd$_2$Ir$_2$O$_7$\cite{Ma538}. 
The present mechanism, which takes the effect of mobile charges into account, might offer an insight into the DW formation in such systems.

\begin{acknowledgments}
The authors thank M. Udagawa for fruitful discussions in the early stage of the present study.   
The modified KPM-LD simulations were carried out at the Supercomputer Center, 
Institute for Solid State Physics, University of Tokyo.
R.O. is supported by the Japan Society for the Promotion of Science 
through a research fellowship for young scientists and the Program for Leading Graduate Schools (ALPS).
K.B. acknowledges support from the LANL Laboratory Directed Research and Development program, project \#20140458ER. 
This research was supported by KAKENHI (No.~24340076), the Strategic Programs for Innovative Research (SPIRE), MEXT, and the Computational Materials Science Initiative (CMSI), Japan. 
\end{acknowledgments}

%\bibliographystyle{apsrev}
%\bibliography{reference}

\end{document}